\documentclass[twocolumn,showpacs,pra,superscriptaddress,amsmath,amssymb]{revtex4}

\usepackage{booktabs}
\usepackage{color}
\usepackage{graphicx}
\usepackage{dcolumn}
\usepackage{bm}
\usepackage{graphicx}

\begin{document}

\title{Photoionization of Xe and Rn from the relativistic random-phase theory}

\author{Chen-Kai Qiao}
\affiliation{College of Physical Science and Technology, Sichuan University,
Chengdu, Sichuan, 610064}
\affiliation{Institute of Atomic and Molecular Physics, Sichuan University,
Chengdu, Sichuan, 610064}

\author{Hsin-Chang Chi}
\affiliation{Department of Physics, National Dong Hwa University, Shoufeng, Hualien, 97401}

\author{Ming-Chien Hsu}
\affiliation{Department of Physics, National Taiwan University, Taipei, 10617}

\author{Xu-Gen Zheng}
\affiliation{Institute of Atomic and Molecular Physics, Sichuan University,
Chengdu, Sichuan, 610064}

\author{Gang Jiang}
\affiliation{Institute of Atomic and Molecular Physics, Sichuan University,
Chengdu, Sichuan, 610064}

\author{Shin-Ted Lin}
\affiliation{College of Physical Science and Technology, Sichuan University,
Chengdu, Sichuan, 610064}

\author{Chang-Jian Tang}
\affiliation{College of Physical Science and Technology, Sichuan University,
Chengdu, Sichuan, 610064}

\author{Keh-Ning Huang}
\email{knhuang1206@gmail.com}
\affiliation{Institute of Atomic and Molecular Physics, Sichuan University,
Chengdu, Sichuan, 610064}
\affiliation{Department of Physics, National Taiwan University, Taipei, 10617}

\date{\today}

\begin{abstract}

Photoionization cross section $\sigma_{n\kappa}$, asymmetry parameter $\beta_{n\kappa}$, and polarization parameters $\xi_{n\kappa}$, $\eta_{n\kappa}$, $\zeta_{n\kappa}$ of Xe and Rn are calculated in the fully relativistic formalism. To deal with the relativistic and correlation effects, we adopt the relativistic random-phase theory with channel couplings among different subshells. Energy ranges for giant \emph{d}-resonance regions are especially considered.

\end{abstract}


\maketitle

\section{Introduction}

Photoionization is of crucial importance in understanding atomic structures as well as in astrophysical modeling. Electron correlations and relativistic effects may be analyzed through the study of photoionization processes \cite{VUV and Soft X-ray Photoionization}. It is of particular interest to study high-Z atoms for relativistic effects \cite{VUV and Soft X-ray Photoionization,Theodosiou,Johnson1,MCRRPA-Theory,MCRRPA-Results0,Deshmukh,MCRRPA-Review}. Direct results arising from relativistic effects are its influences on subshell cross sections, branching ratios, photoelectron angular distributions, and spin polarizations. Previous studies using the relativistic random-phase theory (RRPT) \cite{VUV and Soft X-ray Photoionization,Theodosiou,Johnson1,MCRRPA-Theory,MCRRPA-Results0,Deshmukh,MCRRPA-Review,Huang0,Kutzner}, which has been called the relativistic random-phase approximation (RRPA), also shows that electron-electron correlations play a crucial role in understanding photoionization processes of atomic systems.

Previous studies mainly focus on the cross sections and angular distributions, and less attention was paid to the spin polarizations. The cross sections and angular distributions of Xe and Rn have been carefully studied for both dipole and quadrapole cases. Theoretical works have been done in the Hatree-Fock-Slater theory, nonrelativistic and relativistic random-phase theory, and time-dependent density functional theory \cite{Huang0,Kutzner,Lindau,Kutzner1,Johnson2,Toffoli,Amusia,Govil,Kumar,Kheifets,Saha}.  These are also investigated experimentally in the previous studies \cite{West,Mcllrath,Becker,Andersen,Samson,Suzuki}. However, a complete analysis of photoionization requires spin polarizations of photoelectrons. The observations of spin polarizations of photoelectrons shall provide us sensitive investigations of relativistic effects as well as interchannel couplings \cite{HuangSpin}.

Spin polarizations of photoelectrons from noble gases and other elements have been studied by some authors and compared with experiments \cite{MCRRPA-Review,Huang0,Huang1,Cherepkov1,Cherepkov2,Cherepkov3a,Cherepkov3b,Snell,Khalil,Cherepkov4,Rossi}. Previous studies reveal that the spin polarization generally is an effect of spin-orbit couplings. In this work, we have carried out an analysis on spin polarizations of photoelectrons for heavy noble gas elements Xe and Rn in the RRPT formalism and compared our results with experiment for the Xe atom.

The giant \emph{d}-resonance region is important in photoionization processes of the elements with \emph{d}-shell electrons, where channel couplings and electron correlations play a crucial role in this region. The random phase theory is an effective approach to deal with these channel couplings and electron correlations. It has been demonstrated that the RRPT formalism, when considering core relaxation effects, can give consistent predictions with experiments \cite{Kutzner1,Johnson2}. While other approaches, \emph{e.g.}, Dirac-Hatree-Fock method, fail to explain experimental results in the giant \emph{d}-resonance region. Therefore in this work, we mainly focus our study on the giant \emph{d}-resonance region.

Recently, there are great interests in producing spin-polarized electrons, which are essential for ingredients of spin current in the spintronics field that is efficient in controlling device, with low power consumption and thus promising to act as an effective alternative to conventional electronics \cite{Zutic,Park,Jozwiak,Zhu}. The spin polarizations of the photoelectrons may be affected by phase-shift effects during the optical transition and during the transport through the solid \cite{Heinzmann}. It is thus important to cross-compare results of spin-resolved photoelectrons from the gas phase of free atoms with spin- and angle-resolved photoemission of condensed matter system. Therefore, due to the promising field of spintronics that may become an alternative to electronics by using spin current for several advantages, the understanding of spin polarization is important and has potential in applications nowadays. This work can fill in this gap and give us guidance to the production of spin polarization through circularly polarized, linearly polarized, or even unpolarized light. These can all be parameterized by three dynamical parameters $\xi_{n\kappa}$, $\eta_{n\kappa}$ and $\zeta_{n\kappa}$, which will be defined in Sec. \ref{sec2.2}.

Moreover, liquid detectors based on noble gas elements have been widely applied to detect energetic particles in nuclear and elementary-particle physics. Among them, liquid Xe detectors are predominantly used in recent renewed dark-matter detection and neutrinoless double-beta decay experiments \cite{PandaX1,PandaX2,LUX,XENON1,XENON2,EXO1,EXO2,KamLAND-Zen}. These liquid detectors, which collect signals through ionization processes, are closely related to the photoionization of Xe. Therefore, to better understand the detector response, the interplay between atomic, nuclear, and elementary-particle physics are indispensable. We wish our study on photoionization of Xe and Rn can shed new lights in this area.

\section{Theoretical Treatment}

\subsection{Cross Section}

The basic transition matrix of photoionization process has the form \cite{HuangSpin}

\begin{equation}
T_{fi}=\sqrt{\frac{4\pi^{2}k_{\alpha}E_{\alpha}}{\omega c}}
       \langle\psi_{f}|\sum_{i=1}^{N}\boldsymbol{\alpha}_{i}\cdot\boldsymbol{\hat{\varepsilon}}e^{i\boldsymbol{k}\cdot\boldsymbol{r}_{i}}|\psi_{i}\rangle
\end{equation}
where $\psi_{i}$ and $\psi_{f}$ are the initial and final state respectively, and N is the number of total electrons. We use the boldfaced symbol $\boldsymbol{\alpha}_{i}$ to label the Dirac matrices for $i$-th electron
\begin{equation}
\boldsymbol{\alpha}_{i} = \left(
                          \begin{array}{cc}
                            0                   & \boldsymbol{\sigma}_{i} \\
                            \boldsymbol{\sigma}_{i} & 0
                          \end{array}
                          \right)
\end{equation}
with $\boldsymbol{\sigma}_{i}$ being the Pauli matrices for $i$-th electron. The incident photon has the momentum $\boldsymbol{k}$ and polarization $\boldsymbol{\hat{\varepsilon}}$; the outgoing photoelectron has the momentum $\boldsymbol{k}_{\alpha}$ and energy $E_{\alpha}$, where the subscript $\alpha$, different from the boldfaced symbol $\boldsymbol{\alpha}_{i}$, is the channel index introduced for later convenient. The final state $\psi_{f}$ of the photoelectron and residual ion is normalized such that the differential cross section is given by
\begin{equation}
\frac{d\sigma_{fi}}{d\Omega}=|T_{fi}|^{2}
\end{equation}
The perturbing field can be expanded in a sum of electric and magnetic multipole terms $v_{Jm}^{\lambda}$
\begin{equation}
\sum_{i=1}^{N}\boldsymbol{\alpha}_{i}\cdot\boldsymbol{\hat{\varepsilon}}e^{i\boldsymbol{k}\cdot\boldsymbol{r}_{i}} = \sum_{\lambda Jm}v_{Jm}^{\lambda}
\end{equation}
where the number $J$ corresponds to the $2^{J}$-pole transitions, and $\lambda$ represent the type of transition ($\lambda=E,M$ stands for the electric transition and magnetic transition respectively). Using the Wigner-Eckart theorem, we can decompose the irreducible tensor operator $v_{Jm}^{\lambda}$ into reduced matrix elements
\begin{equation}
\langle u_{\varepsilon_{\alpha}\kappa_{\alpha}} |v_{Jm}^{\lambda}| u_{n\kappa }\rangle
                              = \left(
                                \begin{array}{ccc}
                                  J_{\alpha} & m & m_{0} \\
                                  m_{\alpha} & J & J_{0}
                                \end{array}
                                \right)
                                D_{\alpha}(\lambda J)
\end{equation}
Here we have use the index $\alpha$ to label the ionization channel $\alpha=(n\kappa)\rightarrow(\varepsilon_{\alpha}\kappa_{\alpha})$. The angular momentum of bound state $(n\kappa)$ and continuum state $(\varepsilon_{\alpha}\kappa_{\alpha})$ electron are denote as $J_{0},m_{0}$ and $J_{\alpha},m_{\alpha}$. The symbol $D_{\alpha}(\lambda J)$ denote the reduced amplitudes for electric and magnetic $2^{J}$-pole transitions.

When summing over contributions of transition type as well as all multipole terms $v_{Jm}^{\lambda}$, the total photoionization cross section for state $(n\kappa)$ can be expressed as\cite{HuangSpin}:
\begin{equation}
\sigma_{n\kappa}=\frac{4\pi^{4}c}{\omega(2J_{0}+1)} \bar{\sigma}_{n\kappa}
\end{equation}
where
\begin{eqnarray}
\bar{\sigma}_{n\kappa} & = & \sum_{\lambda JJ_{\alpha}\kappa_{\alpha}}D^{2}_{\alpha}(\lambda J) \nonumber \\
                       & = & \sum_{JJ_{\alpha}\kappa_{\alpha}}[D^{2}_{\alpha}(EJ)+D^{2}_{\alpha}(MJ)]
\end{eqnarray}
The reduced amplitudes $D_{\alpha}(EJ)$ and $D_{\alpha}(MJ)$ for electric $2^{J}$-pole transitions and magnetic $2^{J}$-pole transitions are defined in reference \cite{HuangSpin}. In the electric-dipole approximation, only $D_{\alpha}(E1)$ terms contribute; therefore to represent the reduced dipole amplitudes, we can use a shorthand notation
\begin{equation}
D_{J_{\alpha}} \equiv D_{\alpha}(E1) \label{dipole reduced matrix}
\end{equation}
where $J_{\alpha}$ is the total angular momentum of continuum state electron in channel $\alpha$. Detailed algorithm for computing these reduced amplitudes can be found in references \cite{HuangSpin}.

\subsection{Angular Distribution and Spin Polarization \label{sec2.2}}

Under the electric-dipole approximation, the differential cross section and spin polarization vector $\boldsymbol{P}\equiv\{P_{x},P_{y},P_{z}\}$ of photoelectrons ejected by arbitrarily polarized light are given by \cite{HuangSpin}. We present here only the special cases of circular, linear, and unpolarized light:

$(i)$ circular polarized light
\begin{eqnarray}
\frac{d\sigma_{n\kappa}}{d\Omega} &=& \frac{\sigma_{n\kappa}}{4\pi}
                                      [1-\frac{1}{2}\beta_{n\kappa}P_{2}(\cos{\theta})] \label{differential1a}
\\
P_{x}(\theta,\phi) &=& \frac{\pm\xi_{n\kappa}\sin{\theta}}{1-\frac{1}{2}\beta_{n\kappa}P_{2}(\cos{\theta})}
\\
P_{y}(\theta,\phi) &=& \frac{\eta_{n\kappa}\sin{\theta}\cos{\theta}}{1-\frac{1}{2}\beta_{n\kappa}P_{2}(\cos{\theta})}
\\
P_{z}(\theta,\phi) &=& \frac{\pm\zeta_{n\kappa}\cos{\theta}}{1-\frac{1}{2}\beta_{n\kappa}P_{2}(\cos{\theta})} \label{differential1b}
\end{eqnarray}

$(ii)$ linear polarized light
\begin{eqnarray}
\frac{d\sigma_{n\kappa}}{d\Omega} &=& \frac{\sigma_{n\kappa}}{4\pi}F(\theta,\phi) \label{differential2a}
\\
P_{x}(\theta,\phi) &=& \frac{\eta_{n\kappa}\sin{2\overline{\phi}}\sin{\theta}}{F(\theta,\phi)}
\\
P_{y}(\theta,\phi) &=& \frac{\eta_{n\kappa}\sin{\theta}\cos{\theta}(1+\cos{2\overline{\phi}})}{F(\theta,\phi)}
\\
P_{z}(\theta,\phi) &=& 0
\\
F(\theta,\phi) &=& 1-\frac{1}{2}\beta_{n\kappa}[P_{2}(\cos{\theta})-\frac{3}{2}\cos{2\overline{\phi}}\sin^{2}{\theta}] \nonumber
\\ \label{differential2b}
\end{eqnarray}

\begin{figure}
\includegraphics[width=0.525\textwidth]{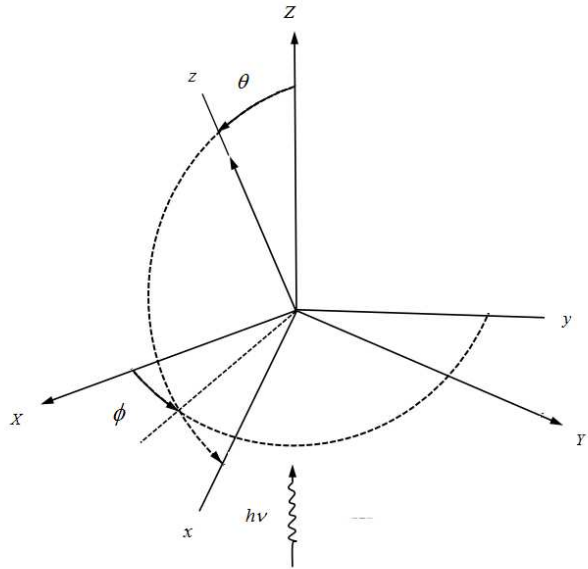}
\caption{ Coordinate systems $XYZ$ and $xyz$. } \label{figure0}
\end{figure}

$(iii)$ unpolarized light
\begin{eqnarray}
\frac{d\sigma_{n\kappa}}{d\Omega} &=& \frac{\sigma_{n\kappa}}{4\pi}[1-\frac{1}{2}\beta_{n\kappa}P_{2}(\cos{\theta})] \label{differential3a}
\\
P_{x}(\theta,\phi) &=& P_{z}(\theta,\phi) = 0
\\
P_{y}(\theta,\phi) &=& \frac{\eta_{n\kappa}\sin{\theta}\cos{\theta}}{1-\frac{1}{2}\beta_{n\kappa}P_{2}(\cos{\theta})} \label{differential3b}
\end{eqnarray}
where $P_{2}(\cos\theta)$ is the Legendre polynomial. The coordinate systems adopted are as follows. The $Z$ axis of the fixed system at the target is chosen along the direction of incoming photon, and the $X$ axis in any convenient direction. The rotated coordinate system $xyz$ are chosen such that $z$ is the direction of photoelectron, therefore the $z$ and $Z$ axes span the reaction plane. The coordinate system $xyz$ is obtained from $XYZ$ through Euler angles $(\phi,\theta,0)$ as shown in Fig.~\ref{figure0}. For linearly polarized light, the angle $\bar{\phi}=\phi-\varphi/2$ in formulas (\ref{differential2a}-\ref{differential2b}) denotes the relative azimuthal angle of photoelectron direction with respect to the linear polarization vector, where the angle $\varphi$ indicates the orientation of linear polarization in the $XY$ plane. The relative azimuthal angle $\bar{\phi}$ depends not only on the direction of photoelectron, but also on the orientation of linear polarization, therefore it is not draw in Fig.~\ref{figure0}. More introduction to angle $\varphi$ and photon spin polarization is given in the Appendix \ref{appendix1}.

From the above equations (\ref{differential1a})-(\ref{differential3b}), it is obvious for angular distribution and spin polarization of photoelectrons, all the angular dependence is contained in trigonometric functions and Legendre polynomial $P_{2}(\cos\theta)$ analytically. The angle-independent dynamical parts, which only depend on the target element and photon energy, are defined as the asymmetry parameter $\beta_{n\kappa}$ and polarization parameters $\xi_{n\kappa}$, $\delta_{n\kappa}$, $\zeta_{n\kappa}$. These parameters, when combined with subshell cross section $\sigma_{n\kappa}$, give a complete description of photoionization processes in the electric-dipole approximation. However, when electric and magnetic multipoles are included, these five parameters $\sigma_{n\kappa}$, $\beta_{n\kappa}$,$\xi_{n\kappa}$, $\delta_{n\kappa}$, $\zeta_{n\kappa}$ are inadequate and we need more dynamical parameters to fully describe the whole photoionization process. More general formulas on angular distributions and spin-polarizations of photoelectrons including electric and magnetic multipole transitions are also shown in Appendix \ref{appendix2}.

The spin-polarization formulas presented above allow us to use dynamical parameters to describe a complete spin-polarization distribution. The spin polarization of the total photoelectron flux is given by $P_{X} = P_{Y} = 0$ and $P_{Z} = \delta_{n\kappa}S_{Z}$, where $\delta_{n\kappa}$ is the polarization parameter defined as
\begin{equation}
\delta_{n\kappa} \equiv \frac{1}{3}(\zeta_{n\kappa}-2\xi_{n\kappa})
\end{equation}

In the electric-dipole approximation, the target electron with initial angular momentum $J_{0}=j$ can be excited to continuum states with angular momenta $J_{\alpha}=j-1$, $j$, and $j+1$. Therefore the photoionization processes are determined by three reduced dipole amplitudes $D_{j-1}$, $D_j$, and $D_{j+1}$ completely. Then the subshell cross section can be expressed as
\begin{equation}
\sigma_{n\kappa} = \frac{4\pi^4 c}{(2J_{0}+1)\omega} \bar{\sigma}_{n\kappa} \label{sigma expression}
\end{equation}
where
\begin{equation}
\bar{\sigma}_{n\kappa} = |D_{j-1}|^2+ |D_j|^2 + |D_{j+1}|^2
\end{equation}
The dynamical parameters for a closed-shell are given by

\begin{eqnarray}
\beta_{n\kappa} & = & \bigg\{
                        \frac{2j-3}{2(2j)}|D_{j-1}|^2
                        -\frac{(2j-1)(2j+3)}{2j(2j+2)}|D_{j}|^2 \nonumber
\\
                &   &   +\frac{2j+5}{2(2j+2)}|D_{j+1}|^2 \nonumber
\\
                &   & -\frac{3}{2j}\bigg[\frac{2j-1}{2(2j+2)}\bigg]^{1/2}(D_{j-1}D_j^*+C.C.) \nonumber
\\
                &   & -\frac{3}{2}\bigg[\frac{(2j-1)(2j+3)}{2j(2j+2)}\bigg]^{1/2}(D_{j-1}D_{j+1}^*+C.C.) \nonumber
\\
                &   & +\frac{3}{2j+2}\bigg[\frac{2j+3}{2(2j)}\bigg]^{1/2}(D_j D_{j+1}^*+C.C.)
                      \bigg\}
                     /\bar{\sigma}_{n\kappa} \nonumber
\\ \label{beta expression}
\end{eqnarray}
\begin{eqnarray}
\xi_{n\kappa} & = & (-)^{l+j+1/2} \nonumber
\\
              &   & \bigg\{
                      -\frac{3(2j-1)}{4(2j)}|D_{j-1}|^2
                      -\frac{3(2j+1)}{2(2j)(2j+2)}|D_{j}|^2 \nonumber
\\
              &   &   +\frac{3(2j+3)}{4(2j+2)}|D_{j+1}|^2 \nonumber
\\
              &   &   +\frac{3}{4(2j)}\bigg[\frac{(2j-1)(2j+2)}{2}\bigg]^{1/2}(D_{j-1}D_j^*+C.C.) \nonumber
\\
              &   &   -\frac{3}{4(2j+2)}\bigg[\frac{2j(2j+3)}{2}\bigg]^{1/2}(D_j D_{j+1}^*+C.C.)
                    \bigg\}
                    /\bar{\sigma}_{n\kappa} \nonumber
\\
\end{eqnarray}
\begin{eqnarray}
\eta_{n\kappa} & = & i(-)^{l+j+1/2} \nonumber
\\
               &   & \bigg\{
                       -\frac{3}{4}\bigg[\frac{2j-1}{2(2j+2)}\bigg]^{1/2}(D_{j-1}D_j^*-C.C.) \nonumber
\\
               &   &   +\frac{3}{4}\bigg[\frac{(2j-1)(2j+3)}{2j(2j+2)}\bigg]^{1/2}(D_{j-1}D_{j+1}^*-C.C.) \nonumber
\\
               &   &   -\frac{3}{4}\bigg[\frac{2j+3}{2(2j)}\bigg]^{1/2}(D_jD_{j+1}^*-C.C.)
                     \bigg\}
                     /\bar{\sigma}_{n\kappa} \nonumber
\\
\end{eqnarray}
\begin{eqnarray}
\zeta_{n\kappa} & = & \bigg\{
                        -\frac{3}{2(2j)}|D_{j-1}|^2
                        +\frac{3}{(2j)(2j+2)}|D_{j}|^2  \nonumber
\\
                &   &   +\frac{3}{2(2j+2)}|D_{j+1}|^2 \nonumber
\\
                &   &   -\frac{3}{2(2j)}\bigg[\frac{(2j-1)(2j+2)}{2}\bigg]^{1/2}(D_{j-1}D_j^*+C.C.) \nonumber
\\
                &   &   -\frac{3}{2(2j+2)}\bigg[\frac{2j(2j+3)}{2}\bigg]^{1/2}(D_j D_{j+1}^*+C.C.)
                      \bigg\}
                      /\bar{\sigma}_{n\kappa} \nonumber
\\ \label{zeta expression}
\end{eqnarray}
where the symbol $C.C.$ in the parentheses stands for complex conjugate. The parameter $\delta_{n\kappa}$ for the total spin polarization is given by

(i) $l=j+1/2$
\begin{eqnarray}
\delta_{n\kappa} & = & \bigg\{
                         \frac{2j-2}{2(2j)}|D_{j-1}|^2
                         +\frac{1}{2j}|D_{j}|^2
                         -\frac{1}{2}|D_{j+1}|^2  \nonumber
\\
                 &   &   -\frac{1}{2j}\bigg[\frac{(2j-1)(2j+2)}{2}\bigg]^{1/2}(D_{j-1}D_j^*+C.C.)
                       \bigg\}
                       /\bar{\sigma}_{n\kappa} \nonumber
\\
\end{eqnarray}

(ii) $l=j-1/2$
\begin{eqnarray}
\delta_{n\kappa} & = & \bigg\{
                         -\frac{1}{2}|D_{j-1}|^2
                         -\frac{1}{2j+2}|D_{j}|^2
                         +\frac{2j+4}{2(2j+2)}|D_{j+1}|^2  \nonumber
\\
                 &   &   -\frac{1}{2j+2}\bigg[\frac{2j(2j+3)}{2}\bigg]^{1/2}(D_jD_{j+1}^*+C.C.)
                       \bigg\}
                       /\bar{\sigma}_{n\kappa} \nonumber
\\ \label{delta expression}
\end{eqnarray}

In addition to subshell cross sections $\sigma_{n\kappa}$ discussed in the previous subsection, the angular distribution and spin polarization can provide rigorous tests for electron correlations. From equations (\ref{beta expression})-(\ref{delta expression}), it is obvious that the relative phase shifts are taken into account in dynamical parameters $\beta_{n\kappa}$, $\xi_{n\kappa}$, $\eta_{n\kappa}$, $\zeta_{n\kappa}$, and $\delta_{n\kappa}$ and therefore can be observed experimentally in angular-distribution and spin-polarization tests.

\subsection{Channels Included \label{sec2.3}}

To analyse channel couplings among different subshells, we have employed a 18-channel calculation in this work, not only consider the allowed open channels in the giant \emph{d}-resonance region, but also include several closed inner-shell channels.

The basic 13 channels allowed in the giant \emph{d}-resonance region we consider are
\[
\begin{array}{ccc}
ns_{1/2}     & \rightarrow & p_{1/2}, p_{3/2}  \\
np_{1/2}     & \rightarrow & s_{1/2}, d_{3/2}  \\
np_{3/2}     & \rightarrow & s_{1/2}, d_{3/2}, d_{5/2} \\
(n-1)d_{3/2} & \rightarrow & p_{1/2}, p_{3/2}, f_{5/2} \\
(n-1)d_{5/2} & \rightarrow & p_{3/2}, f_{5/2}, f_{7/2} \\
\end{array}
\]
where $n=5$ for Xe and $n=6$ for Rn. We also include the following inner-channel contributions:
\[
\begin{array}{ccc}
(n-1)p_{1/2} & \rightarrow & s_{1/2}, d_{3/2}  \\
(n-1)p_{3/2} & \rightarrow & s_{1/2}, d_{3/2}, d_{5/2} \\
\end{array}
\]
These 5 channels, together with the above 13 channels, form the 18 relativistic channels in the context.

\section{Results and Discussions}

\begin{figure*}
\includegraphics[width=1.00\textwidth]{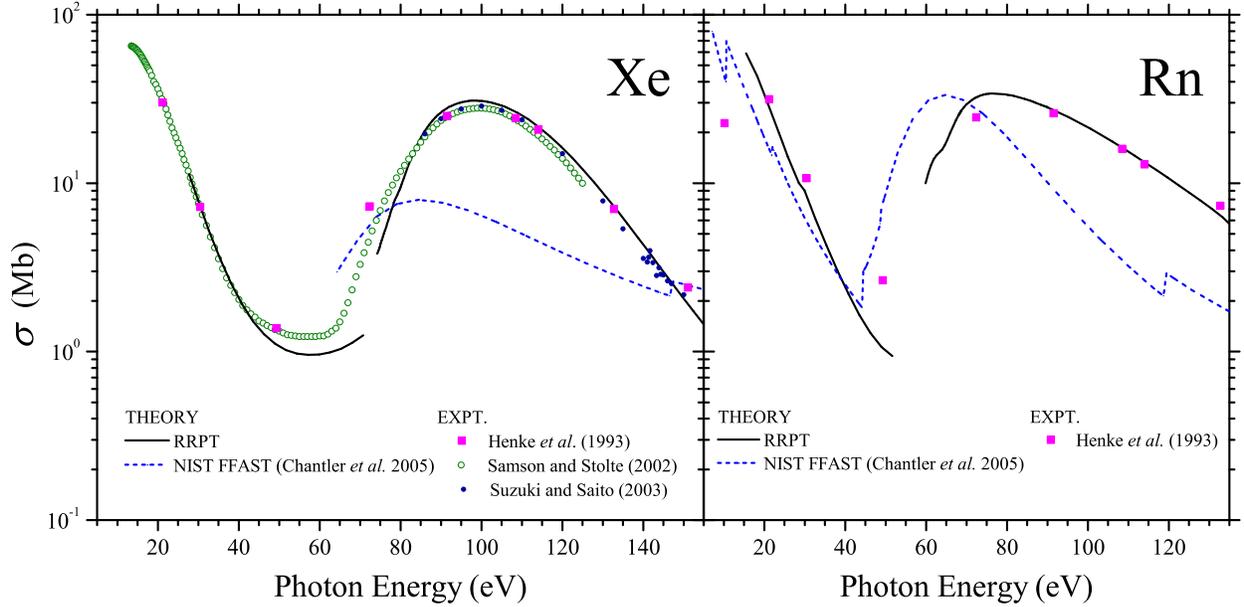}
\caption{Photoionization cross sections of Xe and Rn. The solid lines and dashed lines correspond respectively to our results calculated in RRPT and theoretical results of NIST FFAST database calculated in the Dirac-Hatree-Fock method by Chantler \emph{et al.}. Various experimental results are also plotted in this figure. \label{Total cross section}}
\end{figure*}

\begin{figure*}
\includegraphics[width=1.05\linewidth]{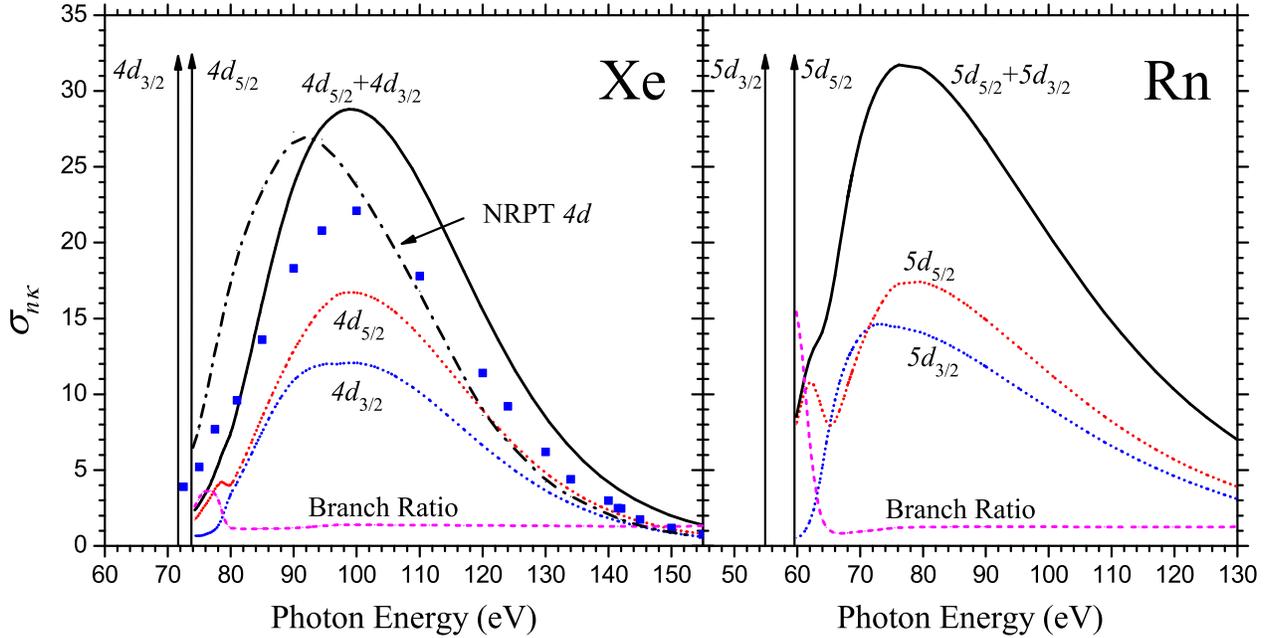}
\caption{Comparison of photoionization cross sections for subshells $(n-1)d_{3/2}$ and $(n-1)d_{5/2}$ of Xe and Rn in the giant \emph{d}-resonance region. The dotted lines correspond to $(n-1)d_{3/2}$ and $(n-1)d_{5/2}$ cross sections from RRPT, where $n=5$ for Xe and $n=6$ for Rn. The squares represent the experimental results for 4\emph{d} subshell of Xe measured by Becker \emph{et al.} \cite{Becker}. Comparison results from nonrelativistic random-phase theory (NRPT) from reference \cite{Kheifets,Saha} are plotted as dashed-dotted line. Branch ratios for subshell cross sections are marked as dashed lines in the figure. The same vertical coordinates are used for subshell cross sections $\sigma_{n\kappa}$ and branch ratios, while cross sections are in units of Mb and branch ratios are dimensionless. The vertical arrows label the photoionization thresholds of $4d$ for Xe and $5d$ for Rn. \label{4d shell cross section}}
\end{figure*}

\begin{figure*}
\includegraphics[width=1.05\textwidth]{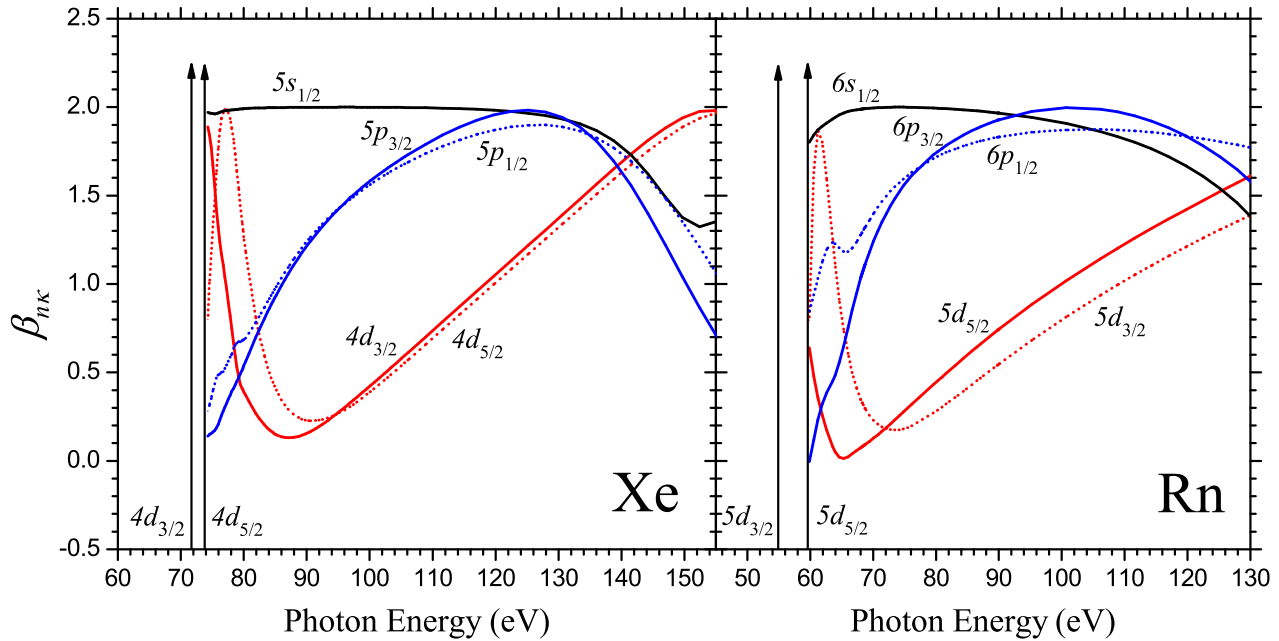}
\caption{Subshell asymmetry parameters $\beta_{n\kappa}$ for Xe and Rn in the giant \emph{d}-resonance region from our RRPT. The vertical arrows label the photoionization thresholds of $4d$ for Xe and $5d$ for Rn.  \label{Figure beta}}
\end{figure*}

In this section, we list our results on photoionization processes of Xe and Rn, carefully analyzing the total cross sections, subsection cross sections, subshell angular distributions and spin polarizations in the giant \emph{d}-resonance region.

The total photoionization cross sections of the Xe and Rn atoms are shown in Fig.~\ref{Total cross section}. Results calculated through the RRPT formalism with 18-channel couplings and those from from NIST FFAST database \cite{Chantler1,Chantler2,NIST FFAST}, which are calculated in a Dirac-Hatree-Fock method by Chantler \emph{et al.}, are displayed in the figure respectively as solid and dashed lines, along with experimental results synthesised by Henke \emph{et al.} \cite{Henke}. In the case of Xe, experimental results from Samson and Stolte \cite{Samson}, Suzuki and Saito are also listed \cite{Suzuki} in the figure. From this figure, it is obvious that RRPT results fit better with experiment than those of NIST FFAST datebase, indicating that random-phase correlations play a dominant role in the giant \emph{d}-resonance region. In the energy region between the $(n-1)d_{3/2}$ and $(n-1)d_{5/2}$ thresholds, 71-74 eV for Xe and 52-59 eV for Rn, there exist many resonance states which could be treated with the quantum-defect theory \cite{Lee,Johnson}, therefore the cross sections in this region are not given in the figure. A detailed study between photoionization thresholds utilizing quantum-defect theory will be undertaken in the near future.

To analyze the channel coupling effects in the giant \emph{d}-resonance region, we plot the subshell cross sections $\sigma_{n\kappa}$ for $(n-1)d_{3/2}$ and $(n-1)d_{5/2}$ in Fig.~\ref{4d shell cross section}, where $n=5$ for Xe and $n=6$ for Rn, and vertical arrows have been used to label the corresponding photoionization thresholds. We compare our results from RRPT calculated with 18-channel couplings and experimental data from Becker \emph{et al.} \cite{Becker}. There are some discrepancies between theory and experiment, which have already seen in previous researches using nonrelativistic and relativistic random-phase theory in the giant \emph{d}-resonance region \cite{Kheifets,Saha}. The nonrelativistic results from reference \cite{Kheifets,Saha} have been compared in this figure. The relativistic results from reference \cite{Saha} are similar to ours, therefore they are not plotted in the figure. These discrepancies between theory and experiment may arise partially from the inadequacy in our calculation to include more inner-shell channel couplings, such as
\[
\begin{array}{ccc}
(n-1)s_{1/2}     & \rightarrow & p_{1/2}, p_{3/2}  \\
(n-1)p_{1/2}     & \rightarrow & s_{1/2}, d_{3/2}  \\
(n-1)p_{3/2}     & \rightarrow & s_{1/2}, d_{3/2}, d_{5/2} \\
(n-2)d_{3/2} & \rightarrow & p_{1/2}, p_{3/2}, f_{5/2} \\
(n-2)d_{5/2} & \rightarrow & p_{3/2}, f_{5/2}, f_{7/2} \\
\end{array}
\]
as well as double-promotion correlations from double excitation channels and correlations from core relaxation. These channels may have resonance effects with our considered 18 channels in subsection \ref{sec2.3}. The branch ratios for Xe and Rn are also depicted in the figure. An interesting phenomenon from the picture is that, subshell cross section $\sigma_{n\kappa}$ for $(n-1)d_{5/2}$, in addition to the giant \emph{d}-resonance, exhibits a small interference peak at lower energy, while $\sigma_{n\kappa}$ for $(n-1)d_{3/2}$ does not show this small resonance, which makes a large variation in the branch ratio in this region. This is mainly due to interchannel couplings mentioned earlier:
\[
\begin{array}{ccc}
(n-1)d_{3/2} & \rightarrow & p_{1/2}, p_{3/2}, f_{5/2} \\
(n-1)d_{5/2} & \rightarrow & p_{3/2}, f_{5/2}, f_{7/2} \\
\end{array}
\]
which was first studied by Amusia \emph{et al.} \cite{Amusia} and later discussed by Kumar \emph{et al.} and Govil \emph{et al.} for Xe and Rn \cite{Kumar,Govil}. These interchannel couplings makes cross sections of $(n-1)d_{5/2}$ appear to have a kink lying 5 eV above the photoionization threshold, while cross sections of $(n-1)d_{3/2}$ are not influenced much by these interchannel couplings. Moreover, the interchannel couplings influence the $(n-1)d_{5/2}$ cross section more for Rn than that of Xe. Previous results from Kumar \emph{et al.} and Govil \emph{et al.} only include several channels in the electric-dipole approximation. Our results, which are calculated in RRPT with 18-channel couplings, demonstrate that these resonant kinks still visible when we include more interchannel couplings

\begin{figure*}
\includegraphics[width=1.05\textwidth]{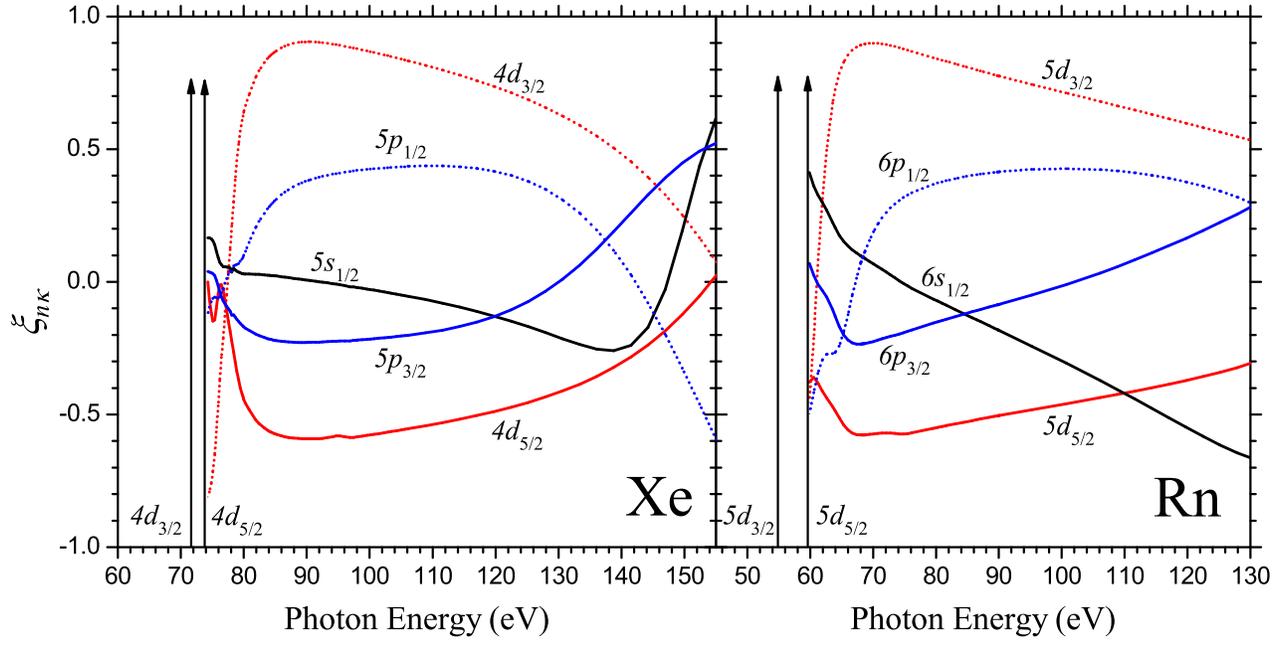}
\caption{Subshell polarization parameters $\xi_{n\kappa}$ for Xe and Rn in the giant \emph{d}-resonance region from our RRPT. The vertical arrows label the photoionization thresholds of $4d$ for Xe and $5d$ for Rn. \label{Figure xi}}
\end{figure*}

\begin{figure*}
\includegraphics[width=1.05\textwidth]{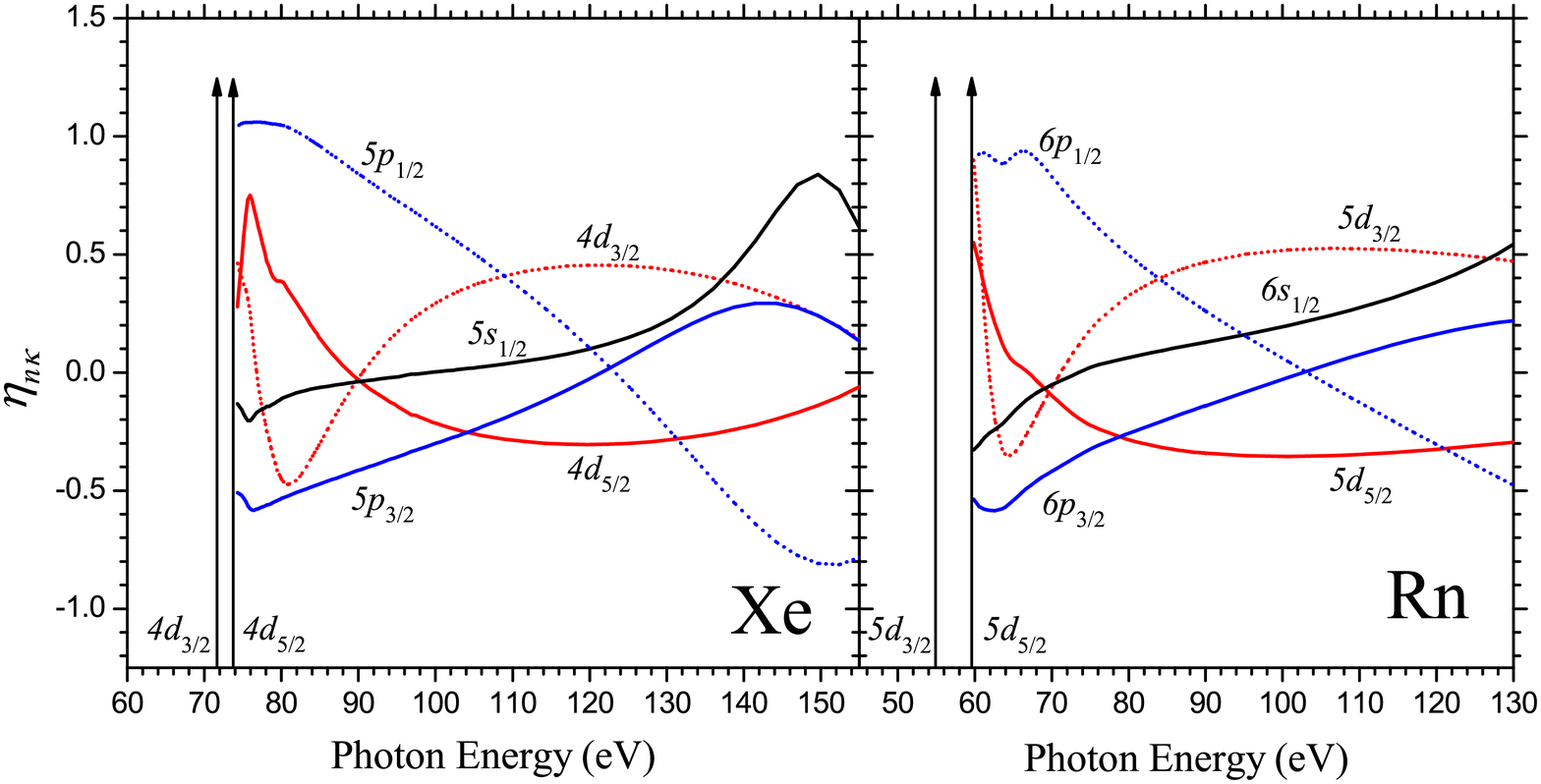}
\caption{Subshell polarization parameters $\eta_{n\kappa}$ for Xe and Rn in the giant \emph{d}-resonance region from our RRPT. The vertical arrows label the photoionization thresholds of $4d$ for Xe and $5d$ for Rn. \label{Figure eta}}
\end{figure*}

\begin{figure*}
\includegraphics[width=1.05\textwidth]{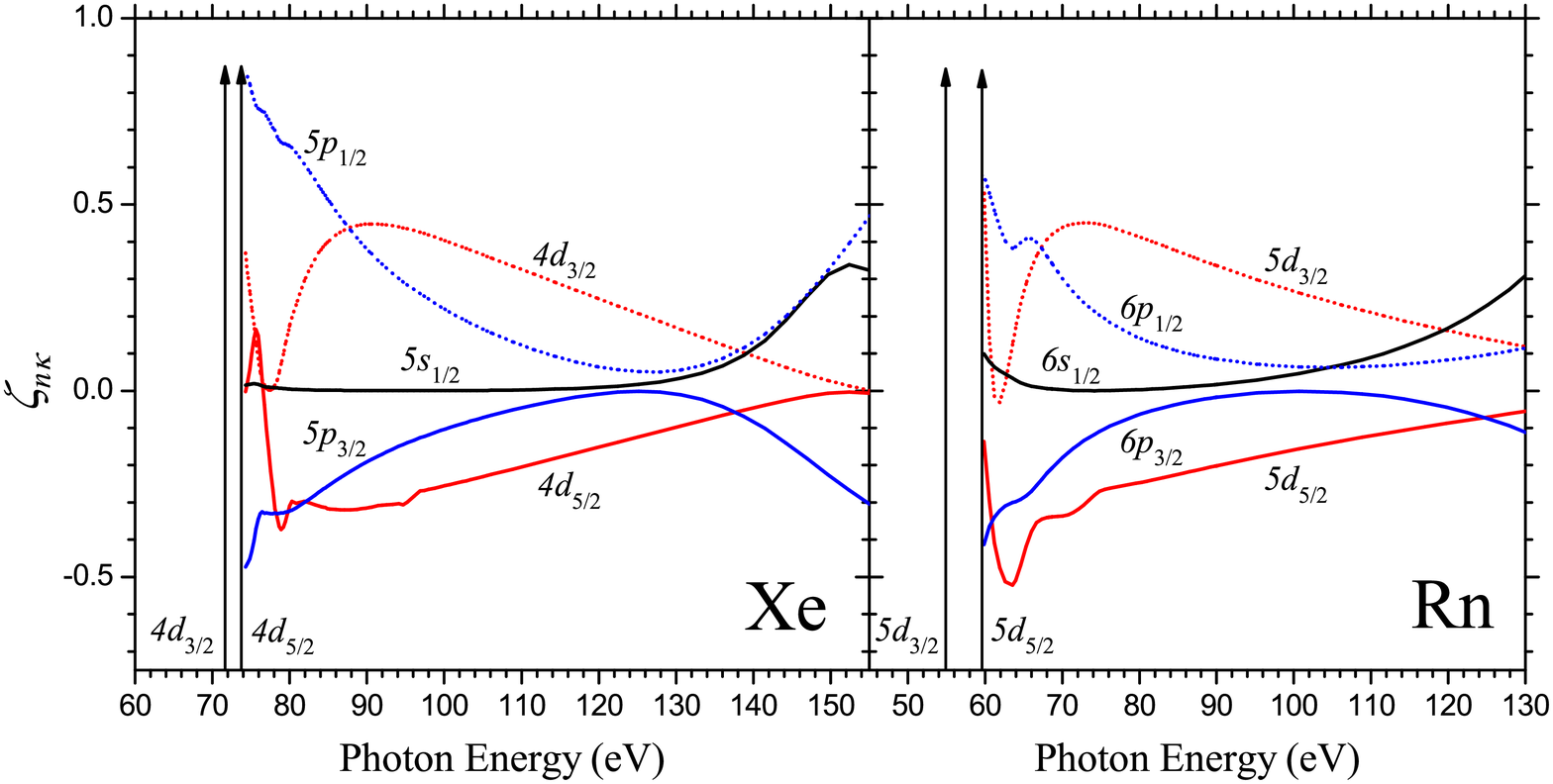}
\caption{Subshell polarization parameters $\zeta_{n\kappa}$ for Xe and Rn in the giant \emph{d}-resonance region from our RRPT. The vertical arrows label the photoionization thresholds of $4d$ for Xe and $5d$ for Rn.  \label{Figure zeta}}
\end{figure*}

\begin{figure*}
\includegraphics[width=1.05\textwidth]{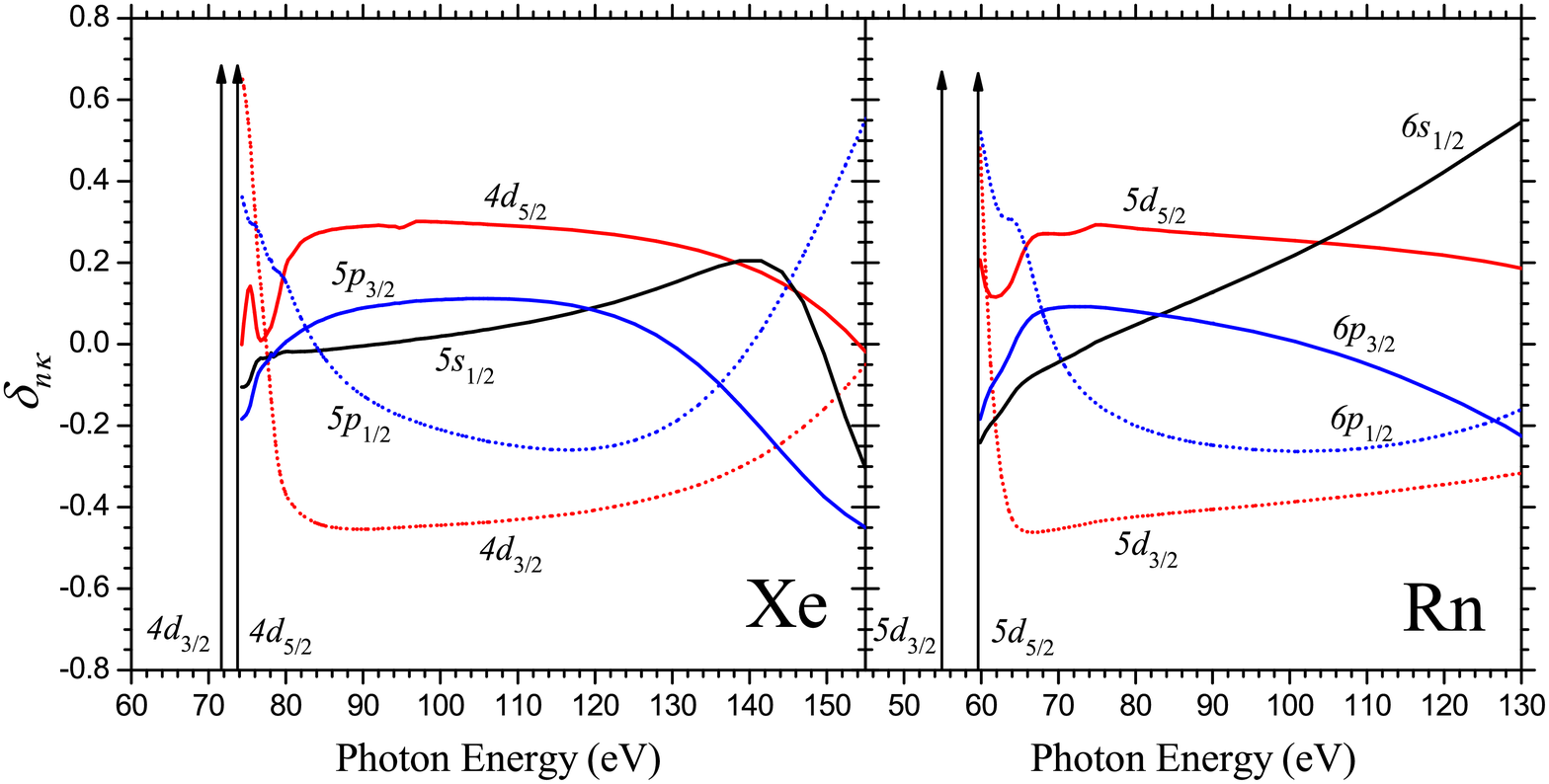}
\caption{Subshell total polarization parameters $\delta_{n\kappa}$ for Xe and Rn in the giant \emph{d}-resonance region from our RRPT. The vertical arrows label the photoionization thresholds of $4d$ for Xe and $5d$ for Rn.  \label{Figure delta}}
\end{figure*}

\begin{figure*}
\includegraphics[width=1.05\textwidth]{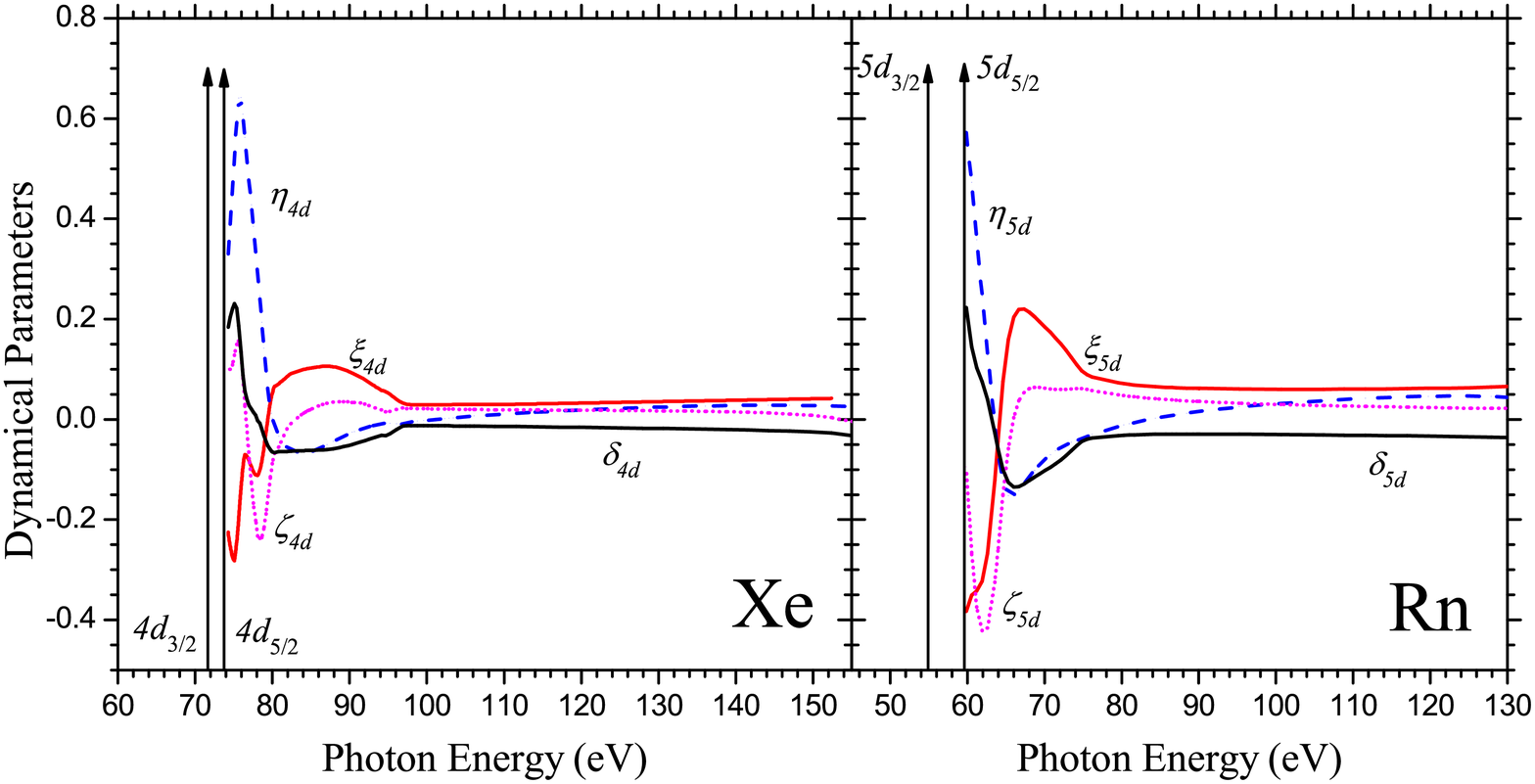}
\caption{Average polarization parameters $\xi_{nl}$, $\eta_{nl}$, $\zeta_{nl}$ and average total polarization parameters $\delta_{nl}$ for closed $(n-1)d$ shell of Xe and Rn in the giant \emph{d}-resonance region calculated in RRPT. The vertical arrows label the photoionization thresholds of $4d$ for Xe and $5d$ for Rn.  \label{Figure D shell}}
\end{figure*}

\begin{figure*}
\includegraphics[width=1.05\textwidth]{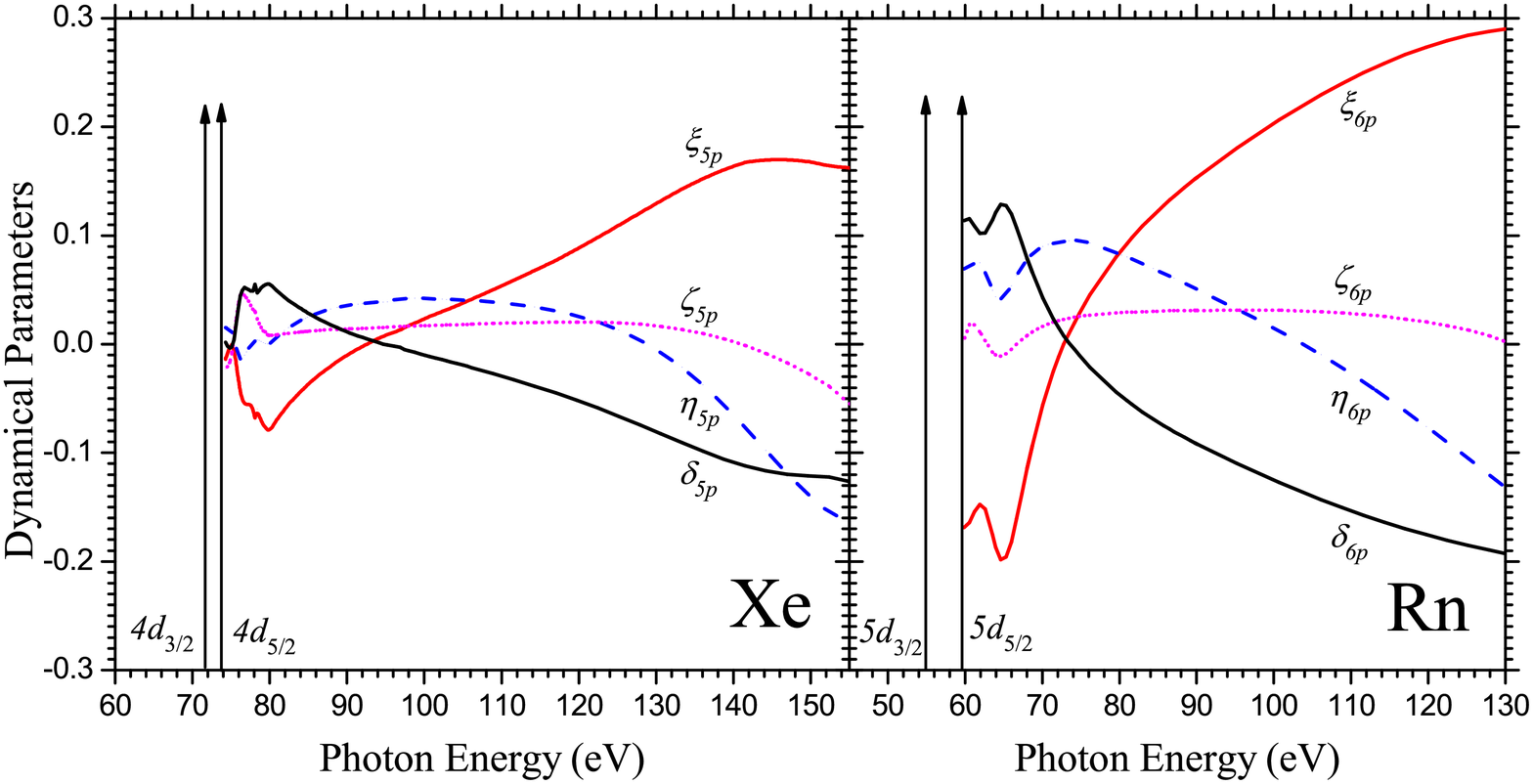}
\caption{Average polarization parameters $\xi_{nl}$, $\eta_{nl}$, $\zeta_{nl}$ and average total polarization parameters $\delta_{nl}$ for closed $np$ shell of Xe and Rn in the giant \emph{d}-resonance region calculated in RRPT. The vertical arrows label the photoionization thresholds of $4d$ for Xe and $5d$ for Rn.  \label{Figure P shell}}
\end{figure*}

The subshell asymmetry parameters $\beta_{n\kappa}$ calculated in RRPT with 18-channel couplings are given in Fig.~\ref{Figure beta}. The vertical arrows are used to label the photoionization thresholds of $(n-1)d_{3/2}$ and $(n-1)d_{5/2}$ as in Fig.~\ref{4d shell cross section}. We find that, in contrast to cross sections, interchannel couplings do not affect much angular distributions of $(n-1)d_{3/2}$ and $(n-1)d_{5/2}$ near the photoionization thresholds, both in cases of Xe and Rn. However, interchannel couplings affect angular distributions of $np_{1/2}$ greatly, giving rise to some structures near photoionization thresholds of $(n-1)d_{3/2}$ and $(n-1)d_{5/2}$.

The subshell polarization parameters $\xi_{n\kappa}$, $\eta_{n\kappa}$, $\zeta_{n\kappa}$ and subshell total polarization parameters $\delta_{n\kappa}$ calculated in RRPT with 18-channel couplings have been displayed in Figs. \ref{Figure xi}-\ref{Figure delta}. In these cases, we find that for parameters $\xi_{n\kappa}$, $\zeta_{n\kappa}$ and $\delta_{n\kappa}$, the interchannel couplings affect spin polarizations of $(n-1)d_{5/2}$, causing the appearance of a kink structure near the photoionization thresholds of $(n-1)d_{3/2}$ and $(n-1)d_{5/2}$, especially in the cases of Xe. In addition, polarization parameters $\eta_{n\kappa}$ and $\zeta_{n\kappa}$ for $np_{1/2}$ also exhibit similar structures near photoionization thresholds, especially in the cases of Rn.

In the nonrelativistic limit, all the polarization parameters are zero for these orbitals $nl$. While in the relativistic cases, due to spin-orbit couplings, the orbitals $nl$ split to orbitals $(nl,l+1/2)$ with parallel spin and orbitals $(nl,l-1/2)$ with antiparallel spin, and all orbitals have nonzero polarization parameters. Moreover, the polarization parameters are different for orbitals with parallel and antiparallel spin, and these are exactly what we have seen in Figs. \ref{Figure xi}-\ref{Figure delta}. Another interesting phenomenon is that, in most of the energy region of giant \emph{d}-resonance, polarization parameters $\xi_{n\kappa}$ and $\zeta_{n\kappa}$ for orbitals with parallel spin, namely $np_{3/2}$ and $(n-1)d_{5/2}$, are less than those with antiparallel spin, $np_{1/2}$ and $(n-1)d_{3/2}$, while polarization parameters $\delta_{n\kappa}$ for orbitals with parallel spin are larger than those with antiparallel spin.

Moreover, one can average over orbitals with parallel and antiparallel spin to get the average polarization parameters $\xi_{nl}$, $\eta_{nl}$, $\zeta_{nl}$ and average total polarization parameter $\delta_{nl}$. For example, we have
\begin{eqnarray}
\delta_{nl} & = & \frac{\sum_{j}\delta_{nlj}\sigma_{nlj}}{\sum_{j}\sigma_{nlj}} \nonumber
                 \\
           & = & \frac{\delta_{nl,l+1/2}\sigma_{nl,l+1/2}+\delta_{nl,l-1/2}\sigma_{nl,l-1/2}}{\sigma_{nl,l+1/2}+\sigma_{nl,l-1/2}}
\end{eqnarray}
In Figs.~\ref{Figure D shell} and \ref{Figure P shell}, we give average polarization parameters $\xi_{nl}$, $\eta_{nl}$, $\zeta_{nl}$ and $\delta_{nl}$ for $(n-1)d$ and $np$ shell of Xe and Rn. In the nonrelativistic limit, the polarization parameters $\xi_{nl}$, $\eta_{nl}$, $\zeta_{nl}$ and $\delta_{nl}$ for closed shells all vanish to zero in the electric-dipole approximation. While in the relativistic case, we get nonzero value of average for all kinds of polarization parameters, as shown in Figs.~\ref{Figure D shell} and \ref{Figure P shell}. For $(n-1)d$ shells, all polarization parameters deviate largely from zero near the ionization threshold, while for $np$ shells, polarization parameters $\xi_{nl}$, $\eta_{nl}$, and $\delta_{nl}$ deviate obviously from zero when photon energy is larger. It is worth noting that, similar to the cases of subshell orbitals, there are also some resonant kink and peak structure for average polarization parameters near the ionization threshold, due to interchannel couplings.

\begin{table}
\caption{Asymmetry parameters $\beta_{n\kappa}$, transferred polarizations $P_{n\kappa}^{\text{transf}}$, and dynamical polarizations $P_{n\kappa}^{\text{dyn}}$ defined by Snell \emph{et al.} \cite{Snell} for each subshell of Xe at energy 93.8 eV. Our present results and experimental measurements from Snell \emph{et al.} are listed here, with numbers in parentheses denoting the experimental uncertainties. For $5p_{1/2}$ and $5p_{3/2}$, the subshell asymmetry parameters $\beta_{n\kappa}$ are not given individually in Snell's work, and the result in table $\beta=1.42_{-0.12}^{+0.08}$ is the average asymmetry parameter of $5p$ closed shell. \label{table}}
\label{table cross section}
\begin{ruledtabular}
\begin{tabular}{lccccc}
Subshell & Parameter             & Present work & Experiment \cite{Snell}
\\
\hline
5s         & $\beta$             & 1.9995       & 1.97(3)
\\
           & $P^{\text{transf}}$ & $-0.0053$    & 0.03(5)
\\
           & $P^{\text{dyn}}$    & 0.0167       & $-0.01(1)$
\\
\hline
$5p_{1/2}$ & $\beta$             & 1.3857       & $1.42_{-0.12}^{+0.08}$
\\
           & $P^{\text{transf}}$ & 0.2993       & 0.31(5)
\\
           & $P^{\text{dyn}}$    & $-0.5609$    & $-0.52(6)$
\\
\hline
$5p_{3/2}$ & $\beta$             & 1.3780       & $1.42_{-0.12}^{+0.08}$
\\
           & $P^{\text{transf}}$ & $-0.1679$    & $-0.18(4)$
\\
           & $P^{\text{dyn}}$    & 0.2765       & 0.31(5)
\\
\hline
$4d_{3/2}$ & $\beta$             & 0.2512       & 0.21(4)
\\
           & $P^{\text{transf}}$ & 0.8448       & 0.83(9)
\\
           & $P^{\text{dyn}}$    & $-0.1232$    & $-0.20(5)$
\\
\hline
$4d_{5/2}$ & $\beta$             & 0.2425       & 0.23(4)
\\
           & $P^{\text{transf}}$ & $-0.5522$    & $-0.57(7)$
\\
           & $P^{\text{dyn}}$    & 0.1118       & 0.15(5)
\end{tabular}
\end{ruledtabular}
\end{table}

To justify our results on angular distributions and spin polarizations in the giant \emph{d}-resonance region, we compare our results with the experimental work of Snell \emph{et al.} \cite{Snell}. Snell \emph{et al.} measure the spin polarization of Xe using spin-resolved photoelectron spectroscopy for a specific geometry $\theta=90^{o}$ and $\phi=135^{o}$ at photon energy 93.8 eV. They have measured asymmetry parameters $\beta_{n\kappa}$, transferred polarizations $P_{n\kappa}^{\text{transf}}$, and dynamical polarizations $P_{n\kappa}^{\text{dyn}}$ for each subshells. Following the definition in \cite{Snell}, for circularly polarized light with Stokes parameters being $S_{X}=S_{Y}=0$, $S_{Z}=\pm1$, the transferred polarization $P_{n\kappa}^{\text{transf}}$ is defined as:
\begin{equation}
P_{n\kappa}^{\text{transf}} \equiv \frac{P_{x}(\theta=90^{o},\phi=135^{o})}{S_{Z}}
                               =   \frac{\xi_{n\kappa}}{1+\beta_{n\kappa}/4}
\end{equation}
While for linearly polarized light, taking a particular polarized direction such that Stokes parameters $S_{X}=1$, $S_{Y}=S_{Z}=0$, the dynamical polarization $P_{n\kappa}^{\text{dyn}}$ is defined as:
\begin{equation}
P_{n\kappa}^{\text{dyn}} \equiv \frac{P_{x}(\theta=90^{o},\phi=135^{o})}{S_{X}}
                            =   -\frac{\eta_{n\kappa}}{1+\beta_{n\kappa}/4}
\end{equation}
The comparison between our theoretical results and experimental measurements with error bars from Snell \emph{et al.} are shown in Table \ref{table}. For $5p_{1/2}$ and $5p_{3/2}$, the individual subshell asymmetry parameters $\beta_{n\kappa}$ are not given in Snell's work, only the average asymmetry parameter for $5p$ shell $\beta_{5p}$ is measured. From Table \ref{table}, it is evident that our results are highly consistent with the experimental data within error bars for this particular geometry at energy 93.8 eV. The above comparisons manifest the validity of our calculation in the giant \emph{d}-resonance region.

\section{Conclusion}

We have carried out a systematic analysis of photoionization processes for Xe and Rn in the fully relativistic RRPT formalism in this paper. The total cross sections $\sigma$, subshell cross sections $\sigma_{n\kappa}$, asymmetry parameters $\beta_{n\kappa}$ and polarization parameters $\xi_{n\kappa}$, $\eta_{n\kappa}$, $\zeta_{n\kappa}$, $\delta_{n\kappa}$ have been analyzed carefully, especially in the giant \emph{d}-resonance region. Our results of total cross sections $\sigma$ match better with the experimental results from X-ray and synchrotron radiation source than those NIST datebases, but small discrepancies still exist and could be treated in multichannel quantum defects theory. For subshell cross sections $\sigma_{n\kappa}$, the interchannel couplings makes $(n-1)d_{5/2}$ exhibits a resonance kink near the photoionization thresholds, especially in the cases of Rn. For angular distributions, asymmetry parameters of $np_{1/2}$ were affected by channel couplings and exhibit similar structures, especially for Rn. For spin polarizations, interchannels couplings influence $4d_{5/2}$ of Xe and $6s_{1/2}$ of Rn greatly, making the appearance of some structures in polarization parameters near photoionization thresholds. For average polarization parameters $\xi_{n\kappa}$, $\eta_{n\kappa}$, $\zeta_{n\kappa}$ and average total polarization parameters $\delta_{n\kappa}$ of $(n-1)d$ and $np$ shells of Xe and Rn, we get non-zero value even in the electric-dipole approximation. Comparisons between theory and experiment on angular distributions and spin polarizations at energy 93.8 eV are provided to demonstrate the validity of our calculation in the giant \emph{d}-resonance region.

Of all the aspects discussed above, the total cross sections $\sigma$ and subshell cross sections $\sigma_{n\kappa}$ have been widely studied, both theoretically and experimentally. For asymmetry parameters $\beta_{n\kappa}$, only a small group of researches focus on this issue, most of them are studied theoretically. For polarization parameters $\xi_{n\kappa}$, $\eta_{n\kappa}$, $\zeta_{n\kappa}$, and $\delta_{n\kappa}$, our knowledge are extremely insufficient. However, because of the rapid development of spintronics, the spin polarizations will play an increasingly significant role in the near future. We wish that the next-generation experiments will measure the angular distribution as well as spin polarization of photoelectrons and therefore provide a more complete analysis for the photoionization processes.

In this work, the photoelectron angular distributions and spin polarizations are analyzed in the electric-dipole approximation. However, as Cherepkov \emph{et al.} have announced in the nonrelativistic cases, the nondipole effects can have notable contributions on angular distributions \cite{Cherepkov3a,Cherepkov3b,Khalil,Cherepkov4}. Analyses which go beyond electric-dipole approximation in fully relativistic cases still leaving to be an extremely valuable exploration in the future.

\section{ACKNOWLEDGMENTS}

The authors acknowledge useful discussions with Professor Hsiang-Shun Chou. Chenkai Qiao is thankful for the accompany of Xiaopang, who is a smart and lively dog. This work was supported by the National Natural Science Foundation of China under Grants No. 11474209, No. 11474208, No. 11475117, and a Grant from the Ministry of Science and Technology of China (No. 2017....).

\appendix

\section{Photon Spin Polarization and Stokes Parameters \label{appendix1}}

There are several equivalent ways of specifying the spin polarization of a given photon beam, and one method is through the density matrix. Here we will adopt the convention used in reference \cite{Huang2}. First, we define the polarization vector $\hat{\varepsilon}$ of photon through the form
\begin{equation}
\hat{\varepsilon}=\hat{e}_{+1}e^{-i\varphi/2}\cos{\theta/2}+\hat{e}_{-1}e^{i\varphi/2}\sin{\theta/2}
\end{equation}
where $\hat{e}_{+1}$ and $\hat{e}_{-1}$ are the spherical unit vectors in a particular coordinate system $XYZ$ (see Fig. \ref{figure0}) and correspond to the positive and negative helicity states, respectively. The spin-density matrix for complete polarized photon can be expressed as
\begin{equation}
\rho_{\text{pol}} = \frac{1}{2}
                    \left( \begin{array}{cc}
                      1+\cos{\theta} & e^{-i\varphi}\sin{\theta} \\
                      e^{-i\varphi}\sin{\theta} & 1-\cos{\theta}
                    \end{array} \right)
\label{spin density matrix1}
\end{equation}
and the spin-density matrix for unpolarized photon is
\begin{equation}
\rho_{\text{unpol}} = \frac{1}{2}
                      \left( \begin{array}{cc}
                        1 & 0 \\
                        0 & 1
                      \end{array} \right)
\end{equation}
The spin-density matrix for generally polarized photon is
\begin{eqnarray}
\rho & = & p\rho_{\text{pol}}+(1-p)\rho_{\text{unpol}} \nonumber
\\
     & = & \frac{1}{2}
           \left( \begin{array}{cc}
             1+p\cos{\theta} & pe^{-i\varphi}\sin{\theta} \\
             pe^{-i\varphi}\sin{\theta} & 1-p\cos{\theta}
           \end{array} \right)
\end{eqnarray}
Here $p$ is the probability of complete polarization, and therefore refer to as the degree of polarization. The light is complete polarized when $p=1$ and is unpolarized when $p=0$. The parameter $\theta$ indicates the type of polarization. The case $\theta=\pi/2$ denotes linear polarization, and the cases $\theta=0$ and $\theta=\pi$ denote the right and left circular polarization (or positive and negative helicity) respectively. The parameter $\varphi$ indicates the orientation of polarization. Specifically, the precession of the electric field for an elliptically polarized photon trace out an ellipse with its major principle axis, making an angle $\varphi/2$ with the X axis. These parameters $(p,\theta,\varphi)$ are defined with the range: $(0 \leq p \leq 1)$, $0 \leq \theta \leq \pi$ and $0 \leq \varphi \leq 2\pi$.

Another way to describe spin polarization is provided by three Stokes parameters $S_{1}$, $S_{2}$, $S_{3}$, then the spin-density matrix can be expressed as
\begin{eqnarray}
\rho & = & \frac{1}{2}I(1+\bold{S}\cdot\bold{\sigma}) \nonumber
\\
     & = & \frac{1}{2}
           \left( \begin{array}{cc}
             1+S_{3} & S_{1}-iS_{2} \\
             S_{1}+iS_{2} & 1-S_{3}
           \end{array} \right)
\label{spin density matrix2}
\end{eqnarray}
where $\bold{S}=(S_{1}$, $S_{2}$, $S_{3})$ is the Stokes vector and $\bold{\sigma}$ is the conventional Pauli matrices.

In (\ref{spin density matrix1}-\ref{spin density matrix2}), the parameters $(p,\theta,\varphi)$ or $(S_{1},S_{2},S_{3})$ can be determined experimentally by three independent measurements. Comparing the expressions of spin-density matrix (\ref{spin density matrix1}) and (\ref{spin density matrix2}), the parameters $(p,\theta,\varphi)$ and $(S_{1},S_{2},S_{3})$ are related to each other through
\begin{eqnarray}
S_{1} & = & p\sin{\theta}\cos{\varphi} \\
S_{2} & = & p\sin{\theta}\sin{\varphi} \\
S_{3} & = & p\cos{\theta}
\end{eqnarray}
Therefore, $S_{1}=\cos{\varphi}$, $S_{2}=\sin{\varphi}$, $S_{3}=0$ correspond to linearly polarized light, $S_{1}=S_{2}=S_{3}=0$ correspond to unpolarized light, and $S_{1}=S_{2}=0$, $S_{3}=\pm1$ correspond to circularly polarized light. Moreover, in the case of circularly polarized light, $S_{3}=1$ and $S_{3}=-1$ represent right-handed and left-handed, respectively.

\section{General Formulas on Angular Distributions and Spin Polarizations \label{appendix2}}

In this Appendix, we give more general expressions on angular distribution and spin polarization go beyond the electric-dipole approximation discussed in subsection \ref{sec2.2}. The angular distribution and spin polarization vectors of photoelectrons including all multipole transitions are given in references \cite{HuangSpin}:
\begin{equation}
\frac{d\sigma}{d\Omega}(\theta,\phi) = \frac{\sigma}{4\pi} F(\theta,\phi) \label{angular distribution}
\end{equation}
where the angular distribution function is
\begin{equation}
F(\theta,\phi) = 1+\sum_{l\geq1}\beta_{0l}d_{00}^{l}+(S_{X}\cos{2\phi}+S_{Y}\sin{2\phi})\sum_{l\geq2}\beta_{1l}d_{20}^{l}
\end{equation}
and
\begin{eqnarray}
P_{x}(\theta,\phi) & = & \bigg\{
                           S_{Z}\sum_{l\geq1}\xi_{3l}d_{01}^{l}+(S_{X}\sin{2\phi}-S_{Y}\cos{2\phi}) \nonumber
                         \\
                   &   &   \sum_{l\geq2}(\xi_{2l}d_{21}^{l}+\eta_{2l}d_{2,-1}^{l})
                         \bigg\}
                         /F(\theta,\phi) \label{spin polarization x}
\\
P_{y}(\theta,\phi) & = & \bigg\{
                           \sum_{l\geq1}\eta_{0l}d_{01}^{l}+(S_{X}\cos{2\phi}+S_{Y}\sin{2\phi}) \nonumber
                         \\
                   &   &   \sum_{l\geq2}(\xi_{2l}d_{21}^{l}-\eta_{2l}d_{2,-1}^{l})
                         \bigg\}
                         /F(\theta,\phi) \label{spin polarization y}
\end{eqnarray}
\begin{eqnarray}
P_{z}(\theta,\phi) & = & \bigg\{
                           S_{Z}\sum_{l\geq0}\zeta_{3l}d_{00}^{l}+(S_{X}\sin{2\phi}-S_{Y}\cos{2\phi}) \nonumber
                         \\
                   &   &   \sum_{l\geq2}\zeta_{2l}d_{20}^{l})
                         \bigg\}
                         /F(\theta,\phi) \label{spin polarization z}
\end{eqnarray}
Here $d_{mn}^{l}(\theta)$ are the standard $d$ function of the rotation matrices and $d_{00}^{l}(\theta)=P_{l}(\cos\theta)$ is the Legendre polynomial, $S_{X}$, $S_{Y}$, $S_{Z}$ are Stokes parameters. We can see from the expressions (\ref{angular distribution})-(\ref{spin polarization z}) that, besides the total cross section $\sigma$ there are eight kinds of dynamical parameters: $\beta_{0l}$, $\beta_{1l}$, $\xi_{2l}$, $\xi_{3l}$, $\eta_{0l}$, $\eta_{2l}$, $\zeta_{2l}$, $\zeta_{3l}$. Among them, $\beta_{0l}$ and $\beta_{1l}$ are asymmetry parameters related to angular distributions of photoelectrons, while $\xi_{2l}$, $\xi_{3l}$, $\eta_{0l}$, $\eta_{2l}$, $\zeta_{2l}$, and $\zeta_{3l}$ are polarization parameters related to the spin polarizations of photoelectrons in $xyz$ axis.

Moreover, the spin polarization of total photoelectron flux in the coordinate system $XYZ$ are given by
\begin{eqnarray}
P_{X} & = & P_{Y} = 0 \\
P_{Z} & = & \delta_{31}S_{Z}
\end{eqnarray}
where the total polarization parameter is defined as
\begin{eqnarray}
\delta_{31} & \equiv & \frac{1}{3}(\zeta_{31}-\sqrt{2}\xi_{31})
\end{eqnarray}

Under the electric dipole approximation, only eight dynamical parameters survive, i.e., $\sigma$, $\beta_{02}$, $\beta_{12}$, $\xi_{22}$, $\xi_{31}$, $\eta_{02}$, $\eta_{22}$, and $\zeta_{31}$. They reduce to the dynamical parameters $\sigma$, $\beta$, $\xi$, $\eta$, and $\zeta$ in (\ref{sigma expression})-(\ref{zeta expression}) via
\begin{eqnarray}
\sigma     & \rightarrow & \sigma \\
\beta_{02} & \rightarrow & -\frac{1}{2}\beta \\
\beta_{12} & \rightarrow & -\sqrt{\frac{2}{3}}\beta \\
\xi_{22}   & \rightarrow & \eta \\
\xi_{31}   & \rightarrow & \sqrt{2}\xi \\
\eta_{02}  & \rightarrow & \sqrt{\frac{2}{3}}\eta \\
\eta_{22}  & \rightarrow & \eta \\
\zeta_{31} & \rightarrow & \zeta
\end{eqnarray}
Similarly, the total polarization parameter $\delta_{31}$ and $\delta$ are related as
\begin{equation}
\delta_{31} \rightarrow \frac{1}{3}(\zeta-2\xi)=\delta
\end{equation}
Moreover, under electric-dipole approximation, the angular distributions and spin polarizations (\ref{angular distribution})-(\ref{spin polarization z}) in cases of circularly polarized light $S_{X}=S_{Y}=0$, $S_{Z}=\pm1$, linearly polarized light $S_{X}=\cos\varphi$, $S_{Y}=\sin\varphi$, $S_{Z}=0$ and unpolarized light $S_{X}=S_{Y}=S_{Z}=0$ reduce to the formulas (\ref{differential1a})-(\ref{differential3b}) in pervious subsection \ref{sec2.2}. The relative azimuthal angle $\bar{\phi}$ in (\ref{differential2a})-(\ref{differential2b}) can be expressed as $\bar{\phi}=\phi-\varphi/2$


\end{document}